\documentclass[sigconf]{acmart} 
\renewcommand\footnotetextcopyrightpermission[1]{} 
\AtBeginDocument{%
  \providecommand\BibTeX{{%
    \normalfont B\kern-0.5em{\scshape i\kern-0.25em b}\kern-0.8em\TeX}}}

\usepackage[bottom]{footmisc}
\usepackage{romannum}
\usepackage{textcomp}
\usepackage{tabularx}
\usepackage{diagbox}
\usepackage{hhline}

\usepackage{bbding}
\usepackage{pifont}
\usepackage{wasysym}
\usepackage{soul}

\usepackage{tikz}
\usepackage{xcolor}
\newcommand*\circled[1]{\tikz[baseline=(char.base)]{
            \node[shape=circle,fill,inner sep=1pt] (char) {\textcolor{white}{#1}};}}

\usepackage{cleveref}
\crefformat{section}{\S#2#1#3}
\crefformat{subsection}{\S#2#1#3}
\crefformat{subsubsection}{\S#2#1#3}

\usepackage{listings}
\newcommand\YAMLcolonstyle{\color{red}}
\newcommand\YAMLkeystyle{\color{black}\small\ttfamily}
\newcommand\YAMLvaluestyle{\color{blue}}
\makeatletter

\newcommand\language@yaml{yaml}

\expandafter\expandafter\expandafter\lstdefinelanguage
\expandafter{\language@yaml}
{
  keywords={true,false,null,y,n},
  keywordstyle=\color{darkgray}\bfseries,
  basicstyle={\YAMLkeystyle\footnotesize},                                 
  sensitive=false,
  comment=[l]{\#},
  morecomment=[s]{/*}{*/},
  commentstyle=\color{purple},
  stringstyle=\YAMLvaluestyle,
  moredelim=[l][\color{orange}]{\&},
  moredelim=[l][\color{magenta}]{*},
  moredelim=**[il][\YAMLcolonstyle{:}\YAMLvaluestyle]{:},   
  morestring=[b]',
  morestring=[b]",
  literate =    {---}{{\ProcessThreeDashes}}3
                {>}{{\textcolor{red}\textgreater}}1
               {|}{{\textcolor{red}\textbar}}1
                {\ -\ }{{\mdseries\ -\ }}3,
}

\lst@AddToHook{EveryLine}{\ifx\lst@language\language@yaml\YAMLkeystyle\fi}
\makeatother

\newcommand\ProcessThreeDashes{\llap{\color{cyan}\mdseries-{-}-}}

\usepackage{color}
\definecolor{dkgreen}{rgb}{0,0.6,0}
\definecolor{gray}{rgb}{0.5,0.5,0.5}
\definecolor{mauve}{rgb}{0.58,0,0.82}
\usepackage{graphics}
\usepackage{caption}
\usepackage{subcaption}
\graphicspath{{./figures/}}

\begin{document}
\pagenumbering{gobble}

\newcommand{\sys}{FleXR}

\newcommand{\ada}[1]{\textcolor{magenta}{AG: #1}}
\newcommand{\ketan}[1]{\textcolor{blue}{KB: #1}}
\newcommand{\jin}[1]{\textcolor{brown}{JH: #1}}

\newcommand{\red}[1]{\textcolor{red}{#1}}
\newcommand{\green}[1]{\textcolor{green}{#1}}
\newcommand{\blue}[1]{\textcolor{blue}{#1}}

\newcommand{\tlide}{$\sim$}
\newcommand{\mul}{$\times$}
\newcommand{\eg}{\emph{e.g.}}
\newcommand{\ie}{\emph{i.e.}}
\newcommand{\etc}{\emph{etc.}}
\newcommand{\etal}{\emph{et al.}}

\newenvironment{tightitemize}%
 {\begin{list}{$\bullet$}{%
 		\setlength{\leftmargin}{10pt}
        \setlength{\itemsep}{0pt}%
        \setlength{\parsep}{0pt}%
        \setlength{\topsep}{0pt}%
        \setlength{\parskip}{0pt}%
        }%
 }%
{\end{list}}

\title{FleXR: A System Enabling Flexibly Distributed Extended Reality}

\author{Jin Heo}
\affiliation{%
  \institution{Georgia Institute of Technology}
  \city{Atlanta}
  \state{Georgia}
  \country{USA}}
\email{jheo33@gatech.edu}

\author{Ketan Bhardwaj}
\affiliation{%
  \institution{Georgia Institute of Technology}
  \city{Atlanta}
  \state{Georgia}
  \country{USA}}
\email{ketanbj@gatech.edu}

\author{Ada Gavrilovska}
\affiliation{%
  \institution{Georgia Institute of Technology}
  \city{Atlanta}
  \state{Georgia}
  \country{USA}}
\email{ada@cc.gatech.edu}

\renewcommand{\shortauthors}{Jin, et al.}

\begin{abstract}
Extended reality (XR) applications require computationally demanding functionalities with low end-to-end latency and high throughput.
To enable XR on commodity devices, a number of distributed systems solutions enable  offloading of XR workloads on remote servers.
However, they make a priori decisions regarding the offloaded functionalities based on assumptions about operating factors, and their benefits are restricted to specific deployment contexts.
To realize the benefits of offloading in various distributed environments, we present a distributed stream processing system, \sys, which is specialized for real-time and interactive workloads and enables flexible distributions of XR functionalities.
In building \sys, we identified and resolved several issues of presenting XR functionalities as distributed pipelines.
\sys\ provides a framework for flexible distribution of XR pipelines while streamlining development and deployment phases.
We evaluate \sys\ with three XR use cases in four different distribution scenarios.
In the results, the best-case distribution scenario shows up to 50\% less end-to-end latency and 3.9\mul\ pipeline throughput compared to alternatives.
\end{abstract}


\maketitle

\section{Introduction}
\label{sec:intro}
Extended reality (XR), including augmented reality (AR) and virtual reality (VR), involves computationally intensive functionalities such as object detection, localization and mapping, and 3D graphics rendering.
Given the low-latency and high-throughput requirements of XR applications~\cite{morin2022toward, taleb2021extremely}, their associated high computing costs limit the XR experiences that can be supported on resource-constrained mobile devices.

To address this, there have been a number of efforts for distributed XR systems~\cite{lai2019furion, meng2020coterie, liu2020firefly, liu2018cutting, zhang2019rendering, liu2019edge, george2020openrtist, schneider2017augmented, cloudxr, azremoterendering, isar, azcustomvision}.
A common thread across them is to predetermine the functionalities that are offloaded to the server, based on assumptions about environmental factors: the client device capacities, network conditions, and workloads.
They provide support for offloading fixed functional components of perceptions or 
rendering, and their benefits can be realized only in specific deployment contexts.

However, we argue that the distribution contexts will differ vastly in terms of the capabilities of the user devices and offload servers, the connectivity among them, and the XR workloads.
In such scenarios, current solutions will be limited in their ability to provide effective server assistance in various contexts.
Applications would need to rely on combinations of existing techniques to leverage the server in assisting with different functionalities.
It will require significant additional development and configuration efforts.

Currently, flexibility in XR workload distribution is missing due to a lack of adequate systems support.
There are previous offloading systems for flexible function migration~\cite{cuervo2010maui, chun2011clonecloud, kosta2012thinkair},
but it is hard to extend their benefits to XR due to their design limitation of function-level offloading (see \S\ref{sec:related} for more details).
Existing stream processing (SP) libraries can potentially enable flexible workload distribution by creating a pipeline at runtime.
While they are used in use cases similar to XR, \eg, multimedia streaming~\cite{gstreamer} and perception pipelines~\cite{beard2017raftlib, quigley2009ros}, the current SP frameworks lack some of the necessary features to adapt distributed stream processing (DSP) for use in XR.
As described in \S\ref{sec:streamprocessing}, a DSP system for XR should support efficient local communication for collocated pipeline components and blocking and non-blocking communication semantics to express the pipeline dependencies and synchronization.
Moreover, it should provide queue size management and multiple network protocol supports for data freshness requirements of distributed XR pipelines, which are not available in existing solutions.

Simply adding the missing features to existing SP libraries is not sufficient to enable the flexible distribution of XR pipelines.
Even if the SP libraries are extended with those DSP features, there are still issues about how to provide the features properly across the development and deployment phases.
Specifically, in existing SP libraries, a pipeline can be created at runtime, and it requires a user to connect pipeline components (compute kernels) via the developer-specified communication ports.
However, since the communication attributes among compute kernels are determined under the user's pipeline context, the user (not the developer of the kernel) should configure the communication attributes of the connection ports when creating a distributed pipeline.

In response, we present {\bf \sys} -- an open-source, flexibly configurable, and high-performance system for distributed XR.
To bring flexibility, we design \sys\ as a DSP system specialized for XR.
With \sys, XR pipelines can be flexibly created for various distribution scenarios at runtime by a user, without requiring any code modification in pipeline kernels.
We identify the key issues in using DSP systems for XR (\S\ref{sec:spissues}) and describe our design decisions which provide the necessary DSP features and address them (\S\ref{sec:designdecisions}).
\sys\ provides a framework to enable the flexible distribution of XR functionalities, streamlining the development and deployment phases.
The developers write their kernels without considering how and where each kernel runs in user pipelines.
The user can configure the communication attributes of the given components without any change.
This feature is realized by the \sys's kernel design with its port abstractions and interfaces (\S\ref{sec:flexrkernel}).
Once the developer writes a \sys\ compute kernel, 
it can
be flexibly deployed and executed in 
diverse distribution scenarios, per user configuration.

We demonstrate the effectiveness of \sys\ through experimental evaluation with three typical XR use cases and four distribution scenarios (\S\ref{sec:evaluation}).
Compared to the existing distributed XR systems~\cite{chen2018marvel, george2020openrtist, lai2019furion, liu2019edge, liu2018cutting, schneider2017augmented, zhang2019rendering}, \sys\ is shown to support all distribution scenarios by creating distributed pipelines with given kernels at runtime.
Our evaluation results show that the offloading effect of each scenario is different based on the workloads, offloading overheads, and device capacity, which support the importance of flexibility in XR workload distributions.
Overall, this paper makes the following contributions:

\begin{tightitemize}
  \item We describe the limitations of existing distributed XR systems with respect to the need for flexibility, and identify the required features for applying DSP to XR.
  \item We present \sys, a DSP system specialized for distributed XR, which addresses the design issues of DSP for XR and enables flexible distributions of XR pipelines.
    Our evaluation in different distribution scenarios demonstrates that \sys\ practically delivers on the promise of flexibility and performance.
  \item We fully open-source \sys, hoping that it would reduce the barriers for further research in the area of distributed XR\footnote{\url{https://github.com/gt-flexr/FleXR}}.
\end{tightitemize}

\section{Related Work and Motivation}
\label{sec:related}
\noindent\textbf{\bf Previous Work and Limitations. }
Flexibility in XR workload distribution is not currently supported despite 
extensive prior research and commercial solutions to offload XR functionalities on remote servers.
Table~\ref{tab:previoustech} summarizes a number of the existing technologies and what they support to be offloaded.
For VR, graphics operations are usually offloaded for providing realistic experiences via high-quality rendering.
Some studies make use of the characteristics of linear perspective in 3D graphics and split the foreground and background rendering to reuse a hardly changing background~\cite{lai2019furion, meng2020coterie}.
For AR, full and partial offloading of perception modules and graphics rendering have been explored~\cite{schneider2017augmented, liu2019edge, george2020openrtist}.
However, their benefits can be effective only in specific deployment contexts with their static workload distributions.

\begin{table}[t]
  \caption{\label{tab:previoustech} The existing technologies' server-side workloads of their distributed architectures.}
  \resizebox{\linewidth}{!}{
    \begin{tabular}{|l|c|c|c|}
      \hhline{====}
      \diagbox[width=17em]{Technologies}{Server-side Workload}                                          & Full Offloading        & Perceptions           & Application Rendering \\ \hline
      \begin{tabular}[c]{@{}l@{}} Marvel~\cite{chen2018marvel}         \end{tabular}                    &                        & \ding{51}             &                       \\
      \begin{tabular}[c]{@{}l@{}} Glimpse~\cite{chen2015glimpse}       \end{tabular}                    &                        & \ding{51}             &                       \\
      \begin{tabular}[c]{@{}l@{}} OpenRiST~\cite{george2020openrtist}  \end{tabular}                    & \ding{51}              &                       &                       \\
      \begin{tabular}[c]{@{}l@{}} ISAR~\cite{isar}      \end{tabular}                                   & \ding{51}              &                       &                       \\
      \begin{tabular}[c]{@{}l@{}} Furion~\cite{lai2019furion}       \end{tabular}                       &                        &                       & \ding{51}             \\
      \begin{tabular}[c]{@{}l@{}} Liu \emph{et al.}~\cite{liu2019edge} \end{tabular}                    &                        & \ding{51}             &                       \\
      \begin{tabular}[c]{@{}l@{}} Liu \emph{et al.}~\cite{liu2018cutting}      \end{tabular}            &                        &                       & \ding{51}             \\
      \begin{tabular}[c]{@{}l@{}} FireFly~\cite{liu2020firefly}       \end{tabular}                     &                        &                       & \ding{51}             \\
      \begin{tabular}[c]{@{}l@{}} Azure Custom Vision~\cite{azcustomvision}       \end{tabular}         &                        & \ding{51}             &                       \\
      \begin{tabular}[c]{@{}l@{}} Azure Remote Rendering~\cite{azremoterendering}       \end{tabular}   &                        &                       & \ding{51}             \\
      \begin{tabular}[c]{@{}l@{}} Nvidia CloudXR™~\cite{cloudxr}               \end{tabular}            &                        &                       & \ding{51}             \\
      \begin{tabular}[c]{@{}l@{}} Schneider \emph{et al.}~\cite{schneider2017augmented}   \end{tabular} & \ding{51}              &                       &                       \\
      \begin{tabular}[c]{@{}l@{}} Zhang \emph{et al.}~\cite{zhang2019rendering}    \end{tabular}        &                        &                       & \ding{51}             \\
      \hhline{====}
    \end{tabular}
  }
\end{table}

\begin{figure}
  \centering
  \includegraphics[width=\linewidth]{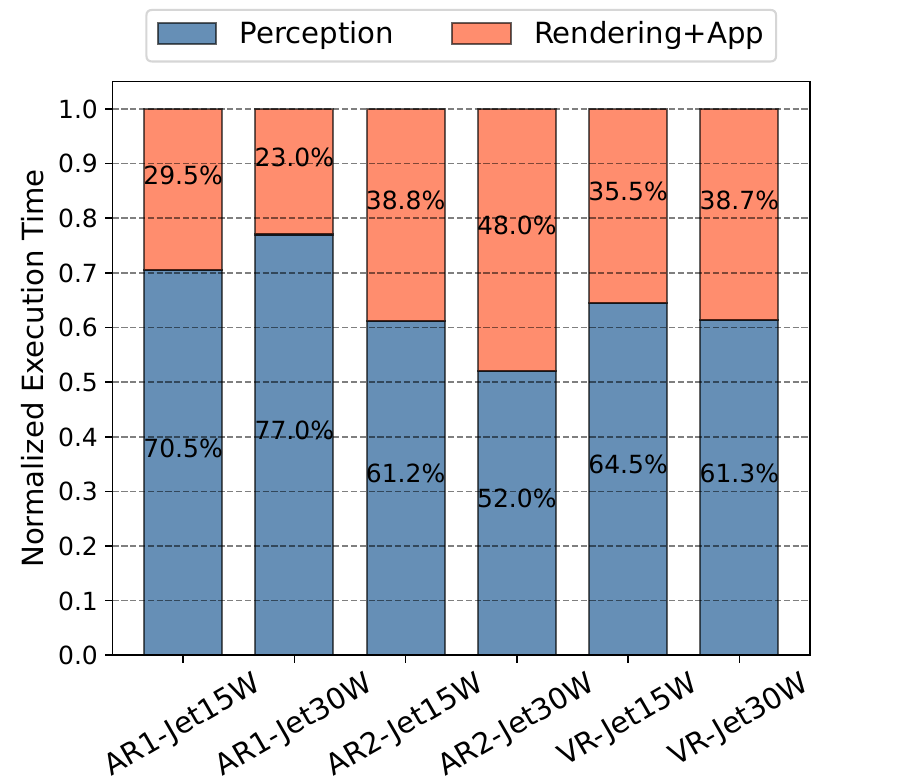}
  \caption{The normalized execution time of our AR and VR examples in different device power modes.}
  \label{fig:workloads}
\end{figure}

Flexibility is necessary because the complexities of XR workloads are not the same for each use case, \eg, AR or VR, and vary based on concrete algorithms and applications.
Figure~\ref{fig:workloads} shows the normalized execution time for the three different AR and VR applications used later in our evaluation.
They are chosen to represent XR use cases with different complexities, which incorporate different algorithms and application frameworks (more detail in \S\ref{sec:exampleapplications} and Figure~\ref{fig:examplefig}).
They run on NVIDIA Jetson AGX~\cite{jetsonagx} with 15W and 30W power modes, which corresponds to the client device in our testbed in \S\ref{sec:evaltestbed}.
These results show each use case has different complexities for its functionalities, and that the dominant functionality -- rendering or perception (or both) -- depends on the workload and the device capacity, making flexibility in offloading an important consideration in distributed XR.

There have been prior works on flexible offloading to a remote server, but they have limited applicability for XR.
MAUI~\cite{cuervo2010maui} proposed fine-grained function offloading with the common language runtime (CLR) of the~.NET framework.
By having CLR on the client and server and requiring developers to specify offloadable functions in application codes, the functions are executed flexibly between the client and server.
CloneCloud~\cite{chun2011clonecloud} and ThinkAir~\cite{kosta2012thinkair} leveraged OS supports to migrate the execution context of threads running application functions to the server's virtual machines (VM);
the server VM provides an environment identical to the client, and the thread execution becomes migratable.

Although these techniques offer some flexibility, their benefit for distributed XR is limited.
In their design approach, devices interact with offloaded functions via client-server interfaces, and the application execution flow is preserved.
Since XR applications require the processing of multimedia data across multiple functionalities, this can introduce multiple network round trips and compromise the benefits of reduced processing time due to offloading.
In addition, preserving the execution flow limits the opportunities to achieve task parallelism.
ThinkAir~\cite{kosta2012thinkair} provides parallelism by cloning VMs, but allows only for data parallelism.
Lastly, since these systems run the application codes on a server with the same environment as the client, they
cannot benefit from 
hardware resources only available on the server.

Offloading XR functionalities requires additional operations such as data compression and network transmissions.
When distributing XR workloads, both overheads from those auxiliary operations and the costs of the XR functionalities should be considered.
Flexibility in reconfiguration thus requires not just techniques which provide transparent function offload, but also systems support to properly configure and deploy the auxiliary functionality.

\noindent\textbf{\bf Motivation for \sys. }
These observations motivate us to build \sys\ based on the stream processing (SP) design.
SP structures an application as a pipeline of components and provides modularity and task parallelism to the application.
A pipeline component (a compute kernel) is the implementation of a functionality.
Kernels are pipelined via data communication ports and executed in parallel with dataflow.
The ports are used to transmit the input and output between the connected kernels.
This modularity makes SP extensible to distributed stream processing (DSP) in a straightforward manner; kernels can be connected via remote communication ports.
Additionally, SP provides an advantage in heterogeneous server environments~\cite{roger2019combining} since kernels can be specialized to utilize available heterogeneous resources such as hardware accelerators.

\section{Design Challenges for \sys}
\label{sec:streamprocessing}
There are intuitive advantages of building an XR system as an SP system.
An XR system processes inputs as data streams from device sensors such as cameras, inertial measurement units, \etc, and provides output as data streams such as field of view content in VR and graphic overlay in AR.
The SP design provides benefits of high throughput with pipeline parallelism and distributed computation with its modularity, but the application of SP in XR presents non-trivial issues.
In this section, we present the design space exploration we conducted to address those issues when building \sys.

\subsection{Issues with Stream Processing for XR}
\label{sec:spissues}
\begin{figure}[t]
  \centering
  \includegraphics[width=\linewidth]{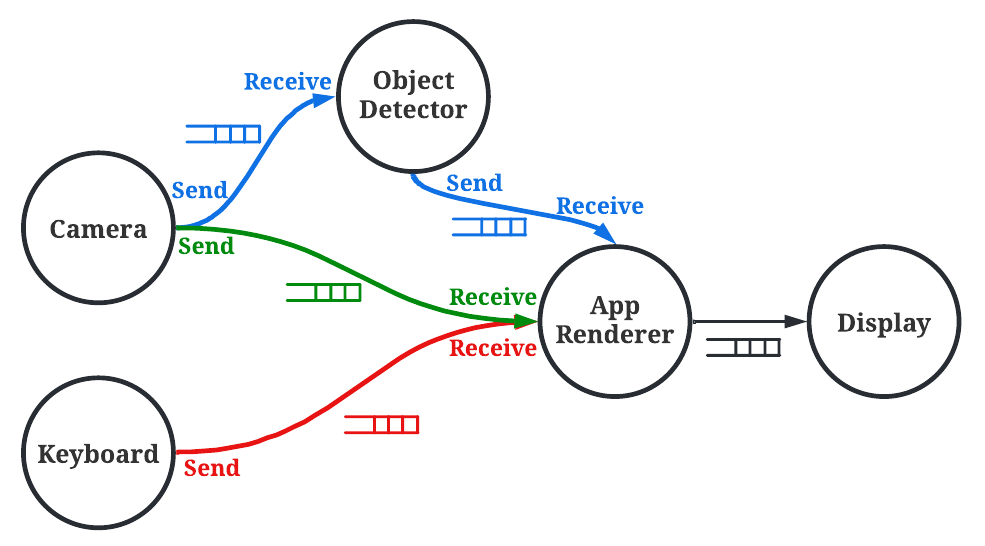}
  \caption{\small The example pipeline of an AR use case.}
  \label{fig:semantics}
\end{figure}

We use an example AR pipeline in Figure~\ref{fig:semantics}, to articulate the issues when applying SP to an XR system.
The application renderer overlays virtual objects on the camera frame based on the result from the object detector.
The objects can be manipulated using key inputs, and the AR scene with the overlaid objects is displayed to the end user.
Using this pipeline, we discuss the main issues below.

\noindent\textbf{\bf I1: Communication Cost. }
As shown in Figure \ref{fig:semantics}, there are a number of components in the pipeline across which data transmissions must occur.
Those data transfers need to be performed with least latency because high latency lowers the application's responsiveness and causes discrepancies between the real and virtual worlds.
However, the SP design increases the end-to-end latency as it requires data movement across the ports of the pipelined kernels~\cite{khare2019linearize, basanta2017patterns}.
Since XR functionalities process and produce large multimedia data, the overhead of the cross-kernel communications is significant and must be addressed in order to meet the latency constraints of XR.

\noindent\textbf{\bf I2: Communication Semantics. }
In Figure~\ref{fig:semantics}, the downstream kernels, \ie, the renderer, has input dependencies with the upstream kernels.
Some of those dependencies are hard, \ie, an input must be received for the downstream kernel to execute, as is the case with the camera and renderer.
Other dependencies are soft, meaning an input is not required from the particular upstream kernel, as is the case with the keyboard or detector and renderer.
This property has implications on whether the execution of the upstream kernels should be synchronized with the downstream kernels in terms of their invocation frequencies.
In short, the type of dependency is specific to the functionalities that are connected, and this semantic information must be expressed by the application and adequately handled by the underlying system via  support for appropriate communication semantics.

\noindent\textbf{\bf I3: Data Recency. }
If the camera frames in Figure~\ref{fig:semantics} are delayed due to queuing delays when the frame data is transmitted, its freshness decreases.
As a result, the placement of the AR object would be off, thus lowering the quality of the AR experience, which is also established by prior work that stale data deteriorates the quality of XR experiences~\cite{li2020towards}.
Generally, as data is transmitted across kernels of different frequencies and execution times, it may result in queuing delays if the data is queued at any of the port buffers in a pipeline.
When the data contains  real-world contexts from sensors, \eg, camera frames, it is critical to ensure that it remains fresh with all pipeline components that process it. Thus, the SP system for XR must provide a way to manage data recency.

\subsection{Design Decisions for The Issues}
\label{sec:designdecisions}
\sys\ as a specialized DSP system for XR, incorporates solutions to address the issues raised in the previous section.

\noindent\textbf{\bf D1: Efficient Local Communication. }
The remote communication cost is unavoidable even with data compression, but the local communication should be efficient for low overheads.
We evaluated the suitability of several existing SP libraries in terms of their communication costs: RaftLib~\cite{beard2017raftlib}, GStreamer~\cite{gstreamer}, Python pipeline libraries~\cite{mpipe, pypeln}, and the robot operating system (ROS)~\cite{quigley2009ros}.

\begin{table}[]
  \caption{\label{tab:localcommunication} \small The local communication latencies between two kernels in milliseconds.}
  \resizebox{0.8\linewidth}{!}{
    \begin{tabular}{|l|c|c|c|c|}
      \hline
      \diagbox{Libraries}{Resolution} & 720p & 1080p & 1440p & 2160p \\ \hline
      ROS Pub/Sub\cite{quigley2009ros}    & 3.4           & 6.9            & 7.1            & 12.5           \\ \hline
      ROS Shm Pub/Sub \cite{rosshm}       & 2.2           & 4.3            & 5.9            & 10.2           \\ \hline
      Python Queue \cite{mpipe, pypeln}   & 14.3          & 24.1           & 30.4           & 52.1           \\ \hline
      Python Pipe \cite{mpipe, pypeln}    & 9.3           & 17.1           & 29.5           & 52.1           \\ \hline
      Python Shm \cite{pythonshm}         & 3.0           & 8.6            & 14.8           & 32.3           \\ \hline
      \textbf{GStreamer} \cite{gstreamer}          & 0.1           & 0.1            & 0.1            & 0.1            \\ \hline
      \textbf{RaftLib} \cite{beard2017raftlib}  & 0.1   & 0.1            & 0.1            & 0.1            \\ \hline
    \end{tabular}
  }
\end{table}

We measure the communication costs with two locally connected kernels and raw RGB frames of different resolutions.
Table~\ref{tab:localcommunication} shows the transmission latencies of the frames.
ROS and Python libraries provide process-level SP, where each kernel runs as a separate process and the processes communicate 
via interprocess communication (IPC) channels.
Based on our results, the local communication in process-level SP is hardly efficient for large multimedia data even with shared memory channels~\cite{pythonshm, rosshm}.
While the shared memory channel reduces the number of data copies, the data still needs to be copied between the shared memory and process memory regions.

GStreamer and RaftLib provide thread-level SP with zero-copy communication ports.
As kernel functions are threads in the same address space, local communication can be done without copy.
\emph{A DSP system for XR should leverage a  thread-level SP for efficient local communication of collocated kernels, and extend its communication with  support for remote communication.}

\noindent\textbf{\bf D2: Blocking and Non-blocking Semantics. }
To handle the hard and soft dependencies and synchronize the kernel executions in a pipeline, providing
the proper communication semantics (blocking and non-blocking) for the local and
remote communication primitives (send and receive) is  essential~\cite{cypher1994semantics}.
The design of \sys\ handles this as a first-order concern when executing XR pipelines.

The send semantics of an output port is for synchronizing the kernel execution.
A blocking send blocks the execution of an upstream kernel function when the downstream kernel's queue is full,
and this backpressure leads to flow control and implicitly synchronizes the upstream to downstream kernels.
For non-blocking semantics, the upstream kernel continues when the downstream kernel cannot receive data on its input port. 
The output port requires both blocking and non-blocking send semantics in XR pipelines.
In the example AR pipeline of Figure~\ref{fig:semantics}, if only blocking semantics are supported, the camera kernel is synchronized to the longest path of object detection (blue line).
Even if the app renderer does not require the results from the object detector for every camera frame, the frame stream (green line) is blocked by the object detector.

The receive semantics of an input port is for kernel dependencies.
A blocking receive waits for the message from a port, and a non-blocking continues when there is no message. 
So, when the kernel is written, the primary inputs on which the kernel 
depends (\eg, camera frame for a kernel performing frame processing) should be specified with blocking semantics.
For inputs generated by other sources (\eg, other sensors or user events), which can impact or steer the kernel processing but are only optionally used, the semantics should be with non-blocking to handle them.

When only a blocking receive is available, all input streams of a kernel are forced as mandatory.
The kernel execution and pipeline throughput are restricted by the lowest frequency input.
In Figure~\ref{fig:semantics}, the renderer is blocked until the key input arrives from the user (red line).
Even without the key input, the renderer execution is governed by the object detector, and the pipeline
throughput is limited by the path with the highest latency (blue line).

\emph{Supporting both blocking and non-blocking primitives for the input and output ports makes it possible
to correctly describe stream dependencies and synchronize kernel executions in XR pipelines.}

\noindent\textbf{\bf D3: Queuing Management and Network Protocols. }
Since poor data freshness causes discrepancies between the real and virtual worlds, it is crucial to manage data recency in XR pipelines.
This can be achieved by minimizing the queuing delays of a pipeline~\cite{little1961proof}.

For local communication, it is possible to bound the queuing delay by limiting the number of outstanding data entries 
in the port buffer.
For remote communication,  recency management becomes challenging because there is no way to control the queuing mechanisms of unknown
middleboxes across the backend network.
In this situation, recency management can be enabled by compromising communication reliability.
Reliable network protocols such as TCP~\cite{stevens1997tcp} guarantee in-order message delivery via retransmission and acknowledgment mechanisms.
However, in cases where recent data is prioritized (\eg, the object detection result on a live camera frame), the reliable protocols are inappropriate, and should be replaced with protocols favoring data timeliness over reliability even with data loss, \eg, RTP~\cite{schulzrinne2003rfc3550} and RTSP~\cite{schulzrinne1998real} over UDP.

Thus, \emph{the DSP system for XR should provide knobs for  data recency management via queue size management and support for multiple network protocols for local and remote kernel communications.}

\section{\sys}
\label{sec:flexr}

\subsection{Overview}
\label{sec:flexroverview}
To bring flexibility to XR workload distribution, we built \sys\ as a DSP system specialized for XR, taking the design benefits of modularity and task parallelism.
Driven by our design decisions, \sys\ is built on top of a thread-level SP library, RaftLib~\cite{beard2017raftlib}, and provides the benefit of efficient local communication for the collocated kernels ({\bf D1}).
For the communication semantics to enable kernel synchronization and dependencies of XR pipelines, we extend the semantics of the RaftLib port with support for non-blocking and for remote communication ({\bf D2}).
For recency management of local and remote communication, \sys\ allows setting the maximum number of messages in the local port buffer and specifying network protocols for remote ports at runtime ({\bf D3}).

Even with the necessary DSP features for XR, there are still issues about providing these features properly to the system stakeholders: developers writing kernels and users requesting distributed pipelines with given kernels.
\sys\ enables flexibility in configuring XR pipelines via its kernel abstraction with interfaces separating development- and deployment-time concerns.
While the developer implements an XR kernel function and knows its input dependencies, the user creates a distributed XR pipeline and configures the connectivity of kernels (local or remote), output-port semantics, and recency management mechanism of the pipeline context.
In addition, there can be a case where the user needs to connect an output of a kernel into multiple downstream kernels.

Our kernel abstraction provides the interfaces allowing the developer to register input and output ports and use the registered ports in the kernel function regardless of how they will be configured by a user.
The behavior of the registered ports becomes different based on the user-specified communication attributes at runtime.
The user can also branch dynamically an output port with different communication attributes and flexibly create distributed pipelines with various topologies without modifying the kernels. 

\begin{figure*}[]
  \centering
  \includegraphics[width=\linewidth]{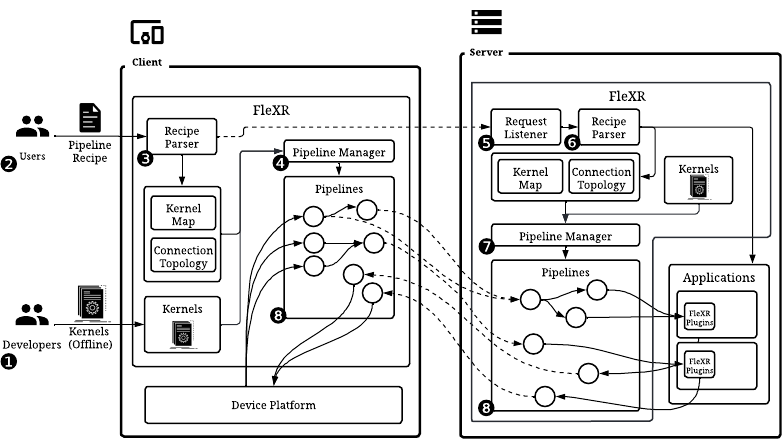}
  \caption{\small The high-level overview of \sys.}
  \label{fig:flexr_overview}
\end{figure*}

\begin{figure}[]
 \centering
  \includegraphics[width=\linewidth]{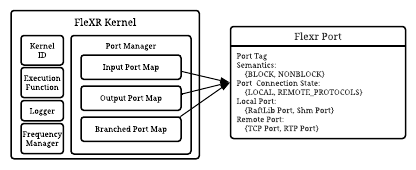}
  \caption{\small \sys\ kernel design with port abstractions.}
  \label{fig:kernel_design}
\end{figure}

Figure~\ref{fig:flexr_overview} shows the high-level design of \sys\ and how it operates.
\circled{1} The developers write kernels with our kernel abstractions for implementing their XR functionalities or incorporating existing functionality implementations by wrapping them in kernel functions.
\circled{2} With given kernels, the user requests a distributed pipeline as a YAML recipe describing pipelined kernels and their communication attributes.
\circled{3} The recipe parser parses the recipe and generates pipeline metadata of the kernels and connection information.
\circled{4} The local pipeline metadata is passed to the pipeline manager, and it creates a local pipeline by instantiating the kernels and configuring port connections.
\circled{5} The part of the recipe about the remote pipeline is sent to the request listener on the server.
\circled{6} The server's recipe parser parses the received recipe and generates pipeline metadata. If the pipeline works with external applications, it starts the applications.
\circled{7} Then, the server's pipeline manager also creates the server-side pipeline, and the remote ports of local and remote pipelines are connected.
\circled{8} The pipelines, distributed across the client and server, run with dataflow.

\subsection{Kernel Design}
\label{sec:flexrkernel}
In SP, the compute kernel is a pipeline component, which includes an execution function and communication ports.
To provide the necessary DSP features for flexible configuration of the communication attributes, we design a \sys\ kernel with two abstractions: the \sys\ port and port manager.
They abstract the different communication channels for local and remote operation of a \sys\ kernel, allowing the developer to write kernels without specifying the communication attributes and the user to configure the connection at runtime without modifying kernels.
A developer registers the input and output ports of the kernel with tags via the port manager interface, and the registered ports are instantiated and configured by the port manager based on the user recipe.

Figure~\ref{fig:kernel_design} shows our kernel design.
Each kernel has its ID, logger, frequency manager, execution function, and port manager.
The ID is used for the recipe parser and pipeline manager in Figure~\ref{fig:flexr_overview}.
The frequency manager adjusts the execution frequency when a kernel should run at a stable frequency, and the logger is for the developer to log the kernel events.
The execution function processes data from the input ports and sends out the result to the output port.

\noindent\textbf{Port Manager.}
When kernels are instantiated, the port manager of each kernel activates and dynamically branches the \sys\ ports based on the pipeline metadata.
In addition, it provides developers with  interfaces to use the \sys\ ports without considering how the ports will be configured by a user.

The port manager design is shown in Figure~\ref{fig:kernel_design}.
The manager has input and output port maps.
These port maps have the mapping information of the port tags registered by a developer and the \sys\ ports activated with the communication attributes by a user.
A kernel function can get inputs and send outputs via the port manager interfaces with the tag.
The branched port map contains the ports branched from the registered output port.
When a registered output port needs to be connected into multiple downstream ports with different communication attributes, the port manager activates the branched ports and keeps their mapping information to the registered port.
When a kernel function sends an output to the registered port, it is also sent through the branched ports by using this mapping information.

Listing~\ref{lst:kernelcode} shows the codes of an example kernel which a developer implements (\circled{1} in Figure~\ref{fig:flexr_overview}).
In Line~\ref{portS}-\ref{portE}, the developer registers the input and output ports with the tags.
The registered ports are used in Line~\ref{getS}-\ref{sendS} without specifying their connection types and branching states.
The port manager hides the complexities of using the dynamically instantiated ports from developers.

\begin{lstlisting}[caption={\small An example kernel with two input and one output ports registered by a developer.}, captionpos=b, label={lst:kernelcode}, float=h]
class ExampleKernel: public FleXRKernel {
  public:
    ExampleKernel() {
      portManager.registerInPortTag("in1", PortSemantics::BLOCKING);(*@\label{portS}@*)
      portManager.registerInPortTag("in2", PortSemantics::NONBLOCKING);(*@\label{portS2}@*)
      portManager.registerOutPortTag("out");(*@\label{portE}@*)
    }

    raft::kstatus run() {
      MsgType *in1=portManager.getInput<MsgType>("in1");(*@\label{getS}@*)
      MsgType *in2=portManager.getInput<MsgType>("in2");(*@\label{getS2}@*)
      MsgType *out=portManager.getOutputPlaceholder<MsgType>("out");(*@\label{getE}@*)

      /* Kernel Functionality ... */

      portManager.sendOutput("out", out);(*@\label{sendS}@*)
    }
}
\end{lstlisting}

\noindent\textbf{\sys\ Port } abstracts different local and remote communication ports and exposes a unified interface to the port manager.
When the pipeline manager creates a pipeline (\circled{4} and \circled{7} in Figure~\ref{fig:flexr_overview}), a kernel is instantiated and its ports are configured by the port manager with user-specified port connectivity, semantics, and recency management mechanism.
Since these communication attributes are determined by the contexts of the requested pipeline, the operation of a \sys\ port should differ based on the attributes given at runtime.
We design the \sys\ port abstraction as a state machine with the integrated interfaces.

The design of a \sys\ port is shown in Figure~\ref{fig:kernel_design}.
Each \sys\ port has the port semantics, connection state, and local and remote ports.
The port semantics is for specifying the communication semantics: blocking and non-blocking.
The connection state indicates whether it is local or remote, and the network protocol for the remote.
The local and remote ports are the actual communication channels internally used and interfaced by the \sys\ port abstraction.
Since the \sys\ port is an abstraction for different communication ports, it is extensible.
New network protocols and local channels can be seamlessly integrated into distributed XR pipelines.

Listing~\ref{lst:yamlrecipe} is part of an example pipeline recipe which a user provides (\circled{2} in Figure~\ref{fig:flexr_overview}).
The user creates a pipeline by specifying the kernels, their port attributes in Line 5, 7-8, 11-12, and 14-15 and connections in Line 17-22.
When the pipeline is created, the port manager activates the \sys\ port.
The activation instantiates an underlying channel corresponding to the specified attributes, and the channel is interfaced via the \sys\ port. 
The \sys\ port provides uniform interfaces to the port manager while
behaving differently based on the underlying channel.

\begin{lstlisting}[caption={\small A part of the pipeline recipe for the example kernel in Listing~\ref{lst:kernelcode} and a connection.}, language=yaml, escapechar=|, captionpos=b, label={lst:yamlrecipe}, float=h]
- kernel   : ExampleKernel
  id       : example_kernel1
  input    :
    - port_name: in1
      connection_type: local
    - port_name: in2
      connection_type: remote
      remote_info: [RTP, 14802]
  output   :
    - port_name: out
      connection_type: local
      semantics: blocking
    - port_name: branched_out
      connection_type: remote
      remote_info    : [127.0.0.1, 14805, TCP]
      branched_from: out
- local_connections:
  - send_kernel: example_kernel1
    send_port_name: out
    recv_kernel: example_kernel2
    recv_port_name: input
    queue_size: 1
\end{lstlisting}

\subsection{Communication Semantics and Data Recency Management}
\label{sec:flexrcommunication}
To express the relationships among kernels and their dependencies and synchronization requirements, both blocking and non-blocking semantics are necessary for the local and remote communication primitives.
\sys\ supports the required semantics.
The local communication in \sys\ is based on the RaftLib port.
Since the send and receive primitives of the vanilla RaftLib port are only with blocking semantics, we extend them with non-blocking semantics by checking the queue buffer of the connected RaftLib ports.
A non-blocking send does not wait and continues when the queue connected to the downstream kernel is full.
A non-blocking receive continues without waiting when the queue to the upstream kernel is empty.
For remote communications, the socket and protocol implementations have interfaces with different semantics, and we map the underlying port interfaces to the \sys\ port.

The data recency management mechanism in \sys\ is to prevent  data from aging in the pipeline queues.
For local, \sys\ provides  recency management by limiting the number of messages in the queue buffer, which puts a bound on the maximum queuing delay~\cite{little1961proof}.
The recency management for remote communication is done by supporting  different network protocols, currently supporting TCP and RTP over UDP.
For TCP connection, the in-order and reliable delivery may lead to lower data timeliness due to its retransmission and acknowledgment mechanisms.
RTP over UDP has the advantage for data recency at the cost of data loss.
By supporting these different protocols and queue size management, the recency management is achieved for remote and local communications.

\subsection{Register-Activation Interface and Port-level Configuration}
\label{sec:flexrportlevel}
We embody the necessary DSP features for XR in the \sys\ kernel, but these features should be provided properly to the stakeholders for supporting the runtime flexibility in distributed XR pipelines.
The kernel developers know the input dependencies of their kernel functions, but it is unknown to them how their kernels are used in a pipeline which a user creates.
When requesting a pipeline, the user arranges the pipeline structure with the kernel communication attributes.
So, the user determines how the kernels operate within the pipeline.
To provide the features to the proper stakeholder, \sys\ has  register-activation interfaces of the port manager at a port granularity, which streamline the development and deployment phases but clearly separate the features provided to each phase.

Based on the information available to the development and deployment phases, we identify the proper stakeholder for each feature and make the interfaces expose it.
Table~\ref{tab:intfeat} summarizes the provided interfaces to each stakeholder.
The developers register ports and set the input-port dependencies as they know the kernel functionalities.
The connection type, branching outstream, output semantics, and recency management are specified by the user recipe because these attributes should be configured when the port manager activates \sys\ ports by the metadata of the user pipeline.

The register-activation interfaces are enabled by the \sys\ port and port manager, and the stakeholders use the \sys\ features through them.
When the developer registers the ports, the semantics of input ports are set via the port manager as shown in Line~\ref{portS}-\ref{portS2} of Listing~\ref{lst:kernelcode}.
The connection types, recency management, and output semantics are specified by the user recipe as shown in Listing~\ref{lst:yamlrecipe}.
The user can branch a single registered port with separate attributes in Line 13-16 of Listing~\ref{lst:yamlrecipe}.
When the pipeline manager instantiates the pipeline kernels, the user-specified attributes and branching are set for each port by the port manager.

\begin{table}[]
  \centering
  \caption{\label{tab:intfeat}\small \sys\ interface availability for the stakeholders to manipulate the features to resolve the DSP issues in \S\ref{sec:streamprocessing}.}
  \resizebox{0.8\linewidth}{!}{
    \tiny
    \begin{tabular}{|l|c|c|}
      \hline
      \diagbox{Feature}{Stakeholder}   & Developer   & User \\ \hline
      Port registration             & \ding{51}   &             \\ \hline
      Port activation               &             & \ding{51}   \\ \hline
      Output branching              &             & \ding{51}   \\ \hline
      Input semantics               & \ding{51}   &             \\ \hline
      Output semantics              &             & \ding{51}   \\ \hline
      Recency management            &             & \ding{51}   \\ \hline
    \end{tabular}
  }
\end{table}

\section{Implementation }
\label{sec:implementation}
The current version of \sys\ is implemented and tested on Ubuntu 20.04.
It is written in C++ with STL, and on-node kernel management and communication rely on RaftLib v0.7~\cite{beard2017raftlib}.
For remote communication, \sys\ supports TCP using ZeroMQ~\cite{zeromq}, and RTP communication using uvgRTP~\cite{altonen2020open} which is based on the  RFC 3550 specifications~\cite{schulzrinne2003rfc3550}.
For the application functionalities, we use several components. For object detectors in the AR applications, we use the ArUco~\cite{romero2018speeded} and ORB keypoint detection algorithm of OpenCV4~\cite{opencv4}.
Pose estimation in the VR application is implemented using ORB SLAM3~\cite{campos2021orb} and the EuRoC dataset~\cite{burri2016euroc}.
\sys\ supports the Unreal Engine 5 (UE5)~\cite{unreal} and Unity 3D~\cite{unity3d} game engines, which interface with the \sys\ runtime via plugins.
The \sys\ plugins are compatible with the shared memory port in Figure~\ref{fig:kernel_design} and make it possible to run external applications with \sys\ pipelines.
The graphics rendering in our examples uses the Mesa implementation~\cite{mesa} of OpenGL~\cite{ogl}, EGL~\cite{egl}, and Vulkan~\cite{vulkan}.
For hardware-accelerated encoding and decoding of H.264~\cite{wiegand2003overview}, we use FFmpeg~\cite{ffmpeg}, NVIDIA Video Codec~\cite{nvcodec} on the server, and NVIDIA L4T~\cite{l4t} on Jetson.

\section{Evaluation}
\label{sec:evaluation}

\begin{figure*}[h!]
  \centering
  \begin{subfigure}[t]{0.2\textwidth}
    \includegraphics[width=\textwidth]{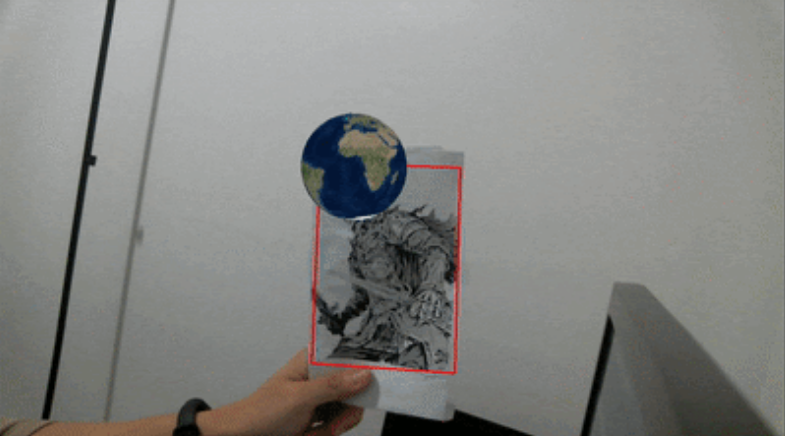}
    \caption{\scriptsize The first AR use case (AR1).}
    \label{fig:ar1}
  \end{subfigure}\hspace{0.1\textwidth}
  \begin{subfigure}[t]{0.2\textwidth}
    \includegraphics[width=\textwidth]{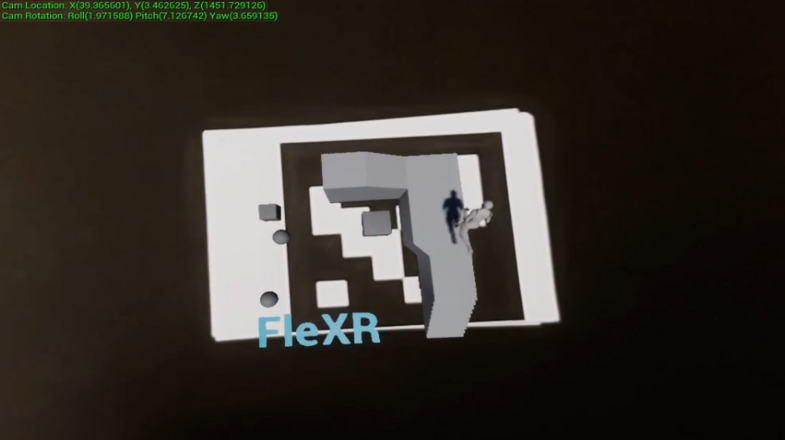}
    \caption{\scriptsize The second AR use case (AR2).}
    \label{fig:ar2}
  \end{subfigure}\hspace{0.1\textwidth}
  \begin{subfigure}[t]{0.2\textwidth}
    \includegraphics[width=\textwidth]{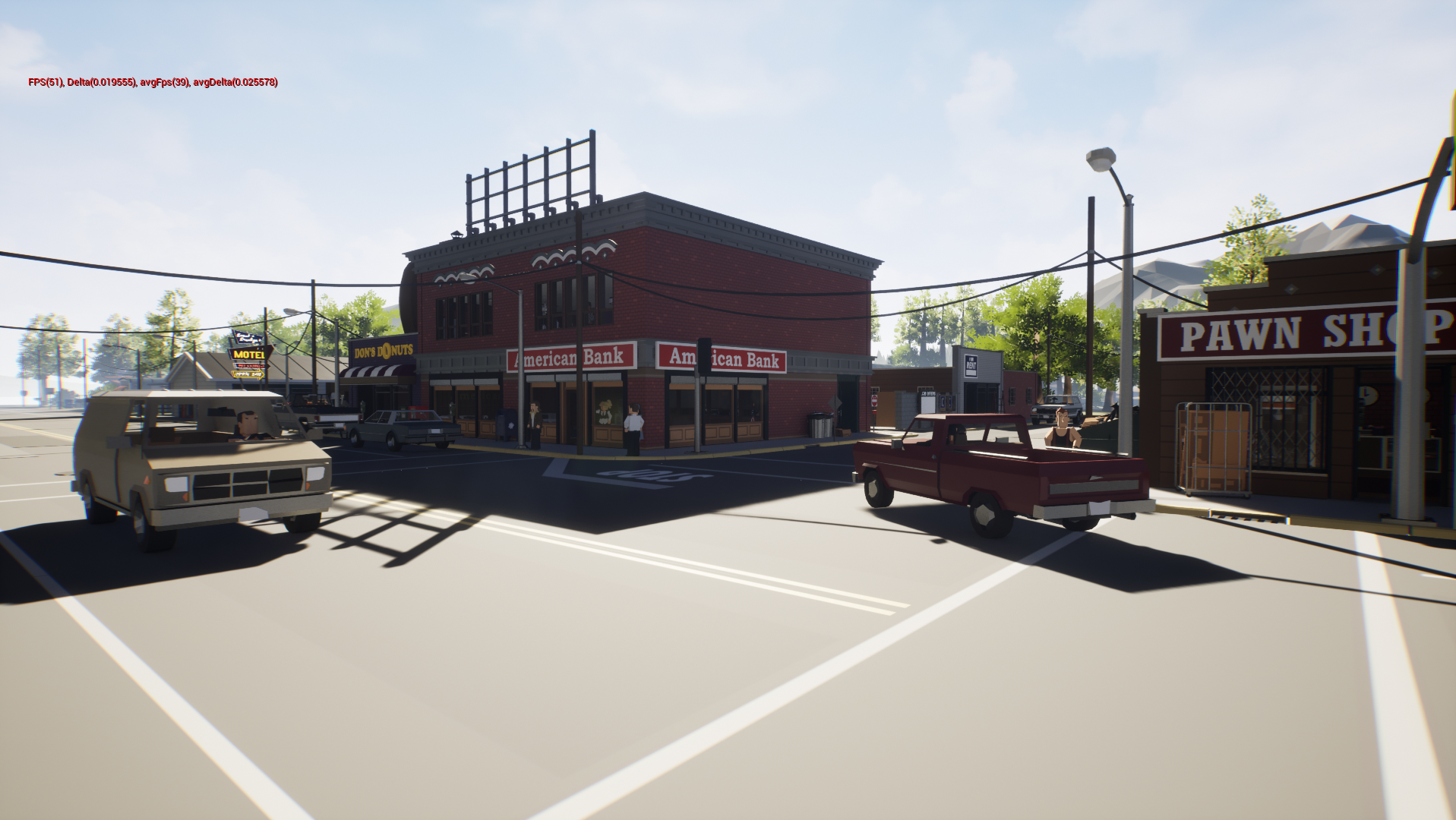}
    \caption{\scriptsize The VR use case (VR).}
    \label{fig:vr}
  \end{subfigure}
  \caption{\small The screenshots of the example use cases.}
  \label{fig:examplefig}
\end{figure*}

\begin{figure*}
  \centering
  \begin{subfigure}[t]{0.2\textwidth}
    \includegraphics[width=\textwidth]{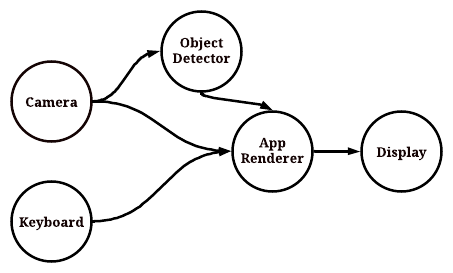}
    \caption{\scriptsize \texttt{Local}: the pipeline runs locally.}
    \label{fig:ar_local}
  \end{subfigure}\hspace{0.05\textwidth}
  \begin{subfigure}[t]{0.2\textwidth}
    \includegraphics[width=\textwidth]{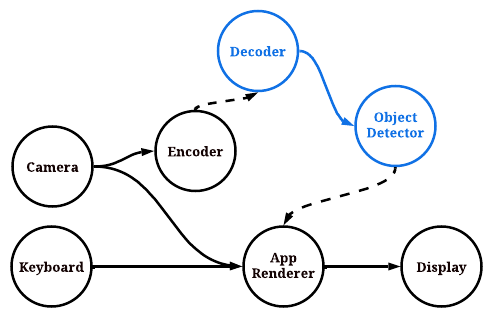}
    \caption{\scriptsize \texttt{Perception}: the perception kernels run on the server.}
    \label{fig:ar_per}
  \end{subfigure}\hspace{0.05\textwidth}
  \begin{subfigure}[t]{0.2\textwidth}
    \includegraphics[width=\textwidth]{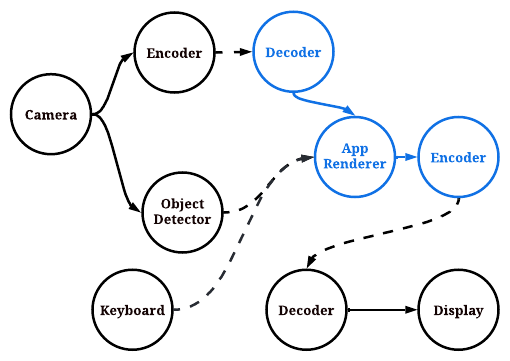}
    \caption{\scriptsize \texttt{Rendering+App}: the application and rendering run on the server.}
    \label{fig:ar_rend}
  \end{subfigure}\hspace{0.05\textwidth}
  \begin{subfigure}[t]{0.2\textwidth}
    \includegraphics[width=\textwidth]{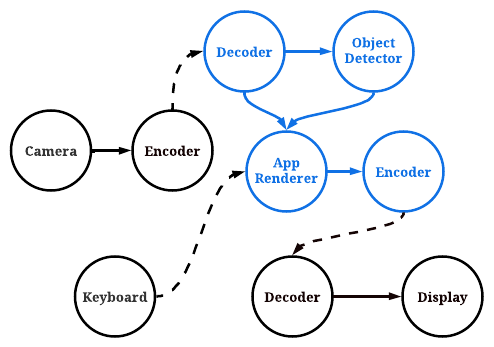}
    \caption{\scriptsize \texttt{Full Offloading}: all functionalities run on the server.}
    \label{fig:ar_full}
  \end{subfigure}
  \caption{\small The distribution scenarios of AR use cases (blue parts on the server).}
  \label{fig:arconfigs}
\end{figure*}

\begin{figure*}
  \centering
  \begin{subfigure}[t]{0.2\textwidth}
    \includegraphics[width=\textwidth]{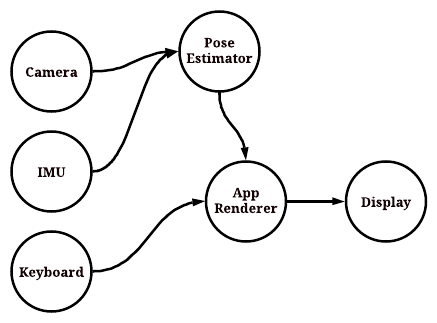}
    \caption{\scriptsize \texttt{Local}: the pipeline runs locally.}
    \label{fig:vr_local}
  \end{subfigure}\hspace{0.05\textwidth}
  \begin{subfigure}[t]{0.2\textwidth}
    \includegraphics[width=\textwidth]{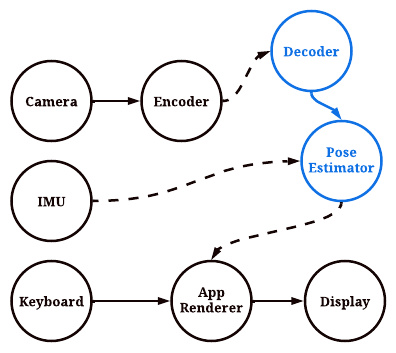}
    \caption{\scriptsize \texttt{Perception}: the perception kernels run on the server.}
    \label{fig:vr_per}
  \end{subfigure}\hspace{0.05\textwidth}
  \begin{subfigure}[t]{0.2\textwidth}
    \includegraphics[width=\textwidth]{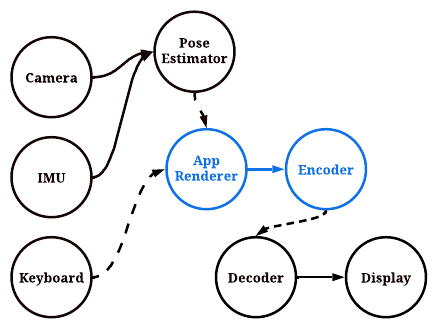}
    \caption{\scriptsize \texttt{Rendering+App}: the application and rendering run on the server.}
    \label{fig:vr_rend}
  \end{subfigure}\hspace{0.05\textwidth}
  \begin{subfigure}[t]{0.2\textwidth}
    \includegraphics[width=\textwidth]{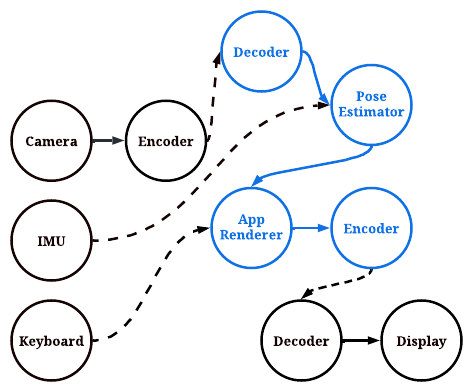}
    \caption{\scriptsize \texttt{Full Offloading}: all functionalities run on the server.}
    \label{fig:vr_full}
  \end{subfigure}
  \caption{\small The distribution scenarios of VR use case (blue parts on the server).}
  \label{fig:vrconfigs}
\end{figure*}

The main objective of \sys\ is to bring flexibility in distributed XR for realizing effective server assistance to XR use cases.
For evaluation, we implement three XR use cases in Figure ~\ref{fig:examplefig} and set four distribution scenarios in Figure~\ref{fig:arconfigs} and~\ref{fig:vrconfigs}.
We compare \sys\ to the existing distributed XR platforms in the supportability to our distribution scenarios.
Then, we evaluate the offloading impacts in the scenarios in terms of pipeline latency and throughput.
Additionally, we evaluate the benefit of the \sys\ design compared to the thread-level SP frameworks: GStreamer~\cite{gstreamer} and RaftLib~\cite{beard2017raftlib}.

\subsection{Experimental Testbed}
\label{sec:evaltestbed}
In our setup, the client is NVIDIA Jetson AGX Xavier~\cite{jetsonagx} with 8 core ARMv8 CPU, Volta GPU, and 32 GB memory shared by CPU and GPU.
The Jetson runs in 15W and 30W power modes (Jet15W using 4 cores and Jet30W using 8 cores).
The server has Intel Core i7-10700, 32 GB memory, and NVIDIA RTX 2070 of 8 GB GDDR6 memory.
The server and client are connected via Gigabit Ethernet of 1 Gbps bandwidth with round-trip time (RTT) of 1.5 ms.

\subsection{XR Applications and Distribution Scenarios}
\label{sec:exampleapplications}
\noindent\textbf{Example XR Applications.}
We evaluate the effectiveness of \sys\ with 2 AR and 1 VR applications as shown in Figure~\ref{fig:examplefig}.
All the applications generate rendered frames of 1080p and provide Full HD (FHD) experiences.
The AR use cases have the same pipeline structure in Figure~\ref{fig:ar_local} taking 1080p camera frames, but with different workload characteristics.
The camera frames are branched to the object detector and renderer because the camera frame needs to be rendered as a background.
The renderer receives the background frame in a blocking manner.
The connection for the object pose detector is non-blocking as an object might not be detected.

For the first AR case (AR1) in Figure~\ref{fig:ar1}, the object detection is done by the local feature matching processes: ORB feature extraction~\cite{rublee2011orb}, k-nearest neighbor (KNN) descriptor matcher, homography and transformation estimations via a perspective-n-point (PnP) random sample consensus (RANSAC) solver~\cite{fischler1981random}.
The second AR case (AR2) in Figure~\ref{fig:ar2} uses the ArUco algorithm~\cite{garrido2014automatic} for detecting the fiducial markers.
While AR2 has a less complex perception than AR1, the application and rendering are more intensive in AR2 as the application is implemented as a separate process by UE5 of the physics and shaders.
On the other hand, AR1 rendering is a pipeline kernel and uses only low-level 3D graphics APIs.

For the VR use case in Figure~\ref{fig:vr}, the pose estimator gets 480p camera frames and inertial measurement unit (IMU) data and generates the current user pose as shown in Figure~\ref{fig:vr_local}.
The pose estimator of the monocular-inertial SLAM has the primary input of IMU and the camera input is optional.
The renderer shows the 3D scene captured from the estimated user pose.
For all example use cases, the user can interact with the virtual objects via keyboard inputs, and this interaction is done by non-blocking receives because the key event happens arbitrarily by the user.

\noindent\textbf{Distribution Scenarios.}
We set up four distribution scenarios for the three use cases: \texttt{Local} (\textbf{L}), \texttt{Perception} (\textbf{P}), \texttt{Rendering+App} (\textbf{R}), and \texttt{Full Offloading} (\textbf{P+R}).
The canonical XR applications consist of perception and graphics rendering functionalities.
As summarized in Table~\ref{tab:previoustech}, the existing distributed XR systems can be categorized into one of our scenarios by their offloading supportability.
With our scenarios, we show the flexibility benefit of \sys\ compared to the existing systems.

\begin{table}[]
  \caption{\label{tab:scenarios} \small The supportability of the
    existing frameworks and \sys\ to our distribution scenarios in
    Figure~\ref{fig:arconfigs} and~\ref{fig:vrconfigs}.}
  \resizebox{\linewidth}{!}{
    \begin{tabular}{|l|c|c|c|c|}
      \hhline{=====}
                                                           & \texttt{Local} & \texttt{Perceptions}   & \texttt{Rendering+App}   & \texttt{Full Offloading} \\ \hline
      Marvel~\cite{chen2018marvel}                         & \ding{55}   & \blue{\ding{51}}     & \ding{55}       & \ding{55}      \\
      OpenRiST~\cite{george2020openrtist}                  & \ding{55}   & \ding{55}     & \ding{55}       & \blue{\ding{51}}      \\
      Furion~\cite{lai2019furion}                          & \ding{55}   & \ding{55}     & \blue{\ding{51}}       & \ding{55}      \\
      Liu \emph{et al.}~\cite{liu2019edge}                 & \ding{55}   & \blue{\ding{51}}     & \ding{55}       & \ding{55}      \\
      Liu \emph{et al.}~\cite{liu2018cutting}              & \ding{55}   & \ding{55}     & \blue{\ding{51}}       & \ding{55}      \\
      Schneider \emph{et al.}~\cite{schneider2017augmented}& \ding{55}   & \ding{55}     & \ding{55}       & \blue{\ding{51}}      \\
      Zhang \emph{et al.}~\cite{zhang2019rendering}        & \ding{55}   & \ding{55}     & \blue{\ding{51}}       & \ding{55}      \\
      \textbf{\sys}                                        & \blue{\ding{51}}   & \blue{\ding{51}}     & \blue{\ding{51}}       & \blue{\ding{51}}      \\ \hhline{=====}
    \end{tabular}
  }
\end{table}

In \texttt{Local} (\textbf{L}), all functionalities run local on the client device only.
In \texttt{Perception} (\textbf{P}), only the perception kernels are offloaded to the server, and in \texttt{Rendering+App} (\textbf{R}) the application rendering are offloaded.
In \texttt{Full Offloading} (\textbf{P+R}), the client only sends the sensor data and receives the final rendered frame.
Figures~\ref{fig:arconfigs} and~\ref{fig:vrconfigs} show the
configurations for the AR1/2 and VR scenarios.
Compared to  existing work which targets specific distributed XR configurations, \sys\ can enable all distribution scenarios flexibly, as shown in Table~\ref{tab:scenarios}.

For the distributed configurations of AR1/2 and VR, the camera and rendered frames and IMUs are transferred with RTP over UDP for application responsiveness while the user input from the keyboard is with TCP for reliable delivery.
When the sensor data is moved via local connections, its queue size is set as 1 to minimize the queuing delay for the data recency.

\begin{figure}[]
  \centering
  \begin{subfigure}[t]{0.2\textwidth}
    \includegraphics[width=\textwidth]{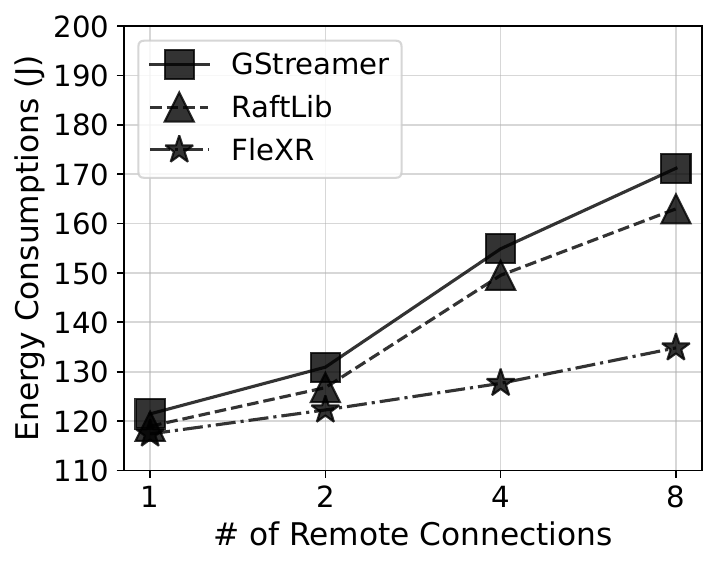}
    \caption{\scriptsize Energy usage results on the server machine.}
    \label{fig:ss2048}
  \end{subfigure}\hspace{0.04\textwidth}
  \begin{subfigure}[t]{0.2\textwidth}
    \includegraphics[width=\textwidth]{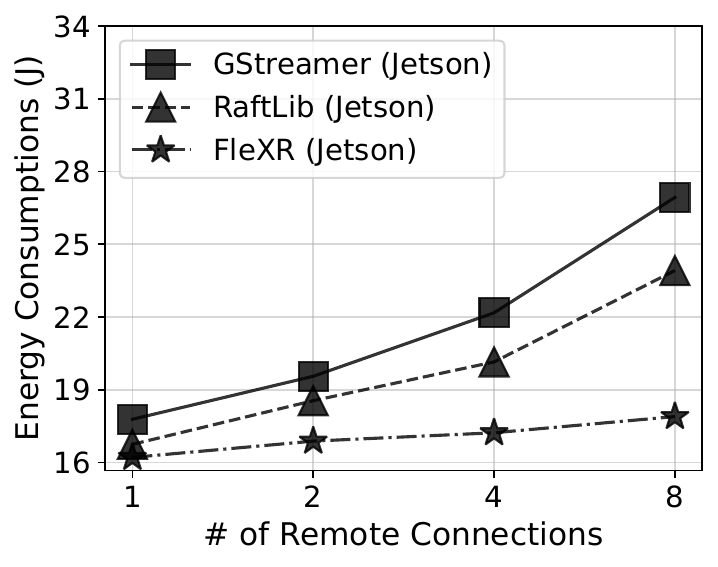}
    \caption{\scriptsize Energy usage results on Jetson 30W.}
    \label{fig:ss1024}
  \end{subfigure}
  \caption {\small Energy consumption to send 1000 messages of 512 Bytes every 10 ms to remote kernels on our server and Jetson 30W.}
  \label{fig:auxcost}
\end{figure}

\subsection{Design Benefit of \sys}
As shown in Table~\ref{tab:localcommunication}, GStreamer and RaftLib provide efficient local communication.
Instead of using \sys, a distributed pipeline can be supported by implementing auxiliary kernels for remote communications, branching, and synchronization.
However, supporting a distributed pipeline with the auxiliary kernels introduces inefficiencies because each kernel, including the auxiliary ones, is a separate execution unit that is parallelly scheduled and managed.

Figure~\ref{fig:auxcost} shows the overheads of the auxiliary kernels on our testbed.
We create a kernel sending output to multiple remote kernels, and measure the energy consumption for the transmissions.
GStreamer and RaftLib need the additional kernels for remote messaging and output branching; for sending output to 8 remote kernels, 9 auxiliary kernels are required (1 for branching and 8 for remote messaging).
The energy consumption with the auxiliary kernels increases with their number.
In contrast, with the \sys\ kernel design and interface, the output port registered by a developer can be branched and configured for remote connections with different protocols as specified by a user, not requiring additional kernels.
In \sys, the developers also don't need to implement these auxiliary kernels to make their kernels operate flexibly.

\begin{table}[b]
  \caption{\label{tab:reqkernels} \small The number of kernels
    required to support the distributed configurations of AR1 in
    Figure~\ref{fig:arconfigs}.}
  \resizebox{\linewidth}{!}{
    \begin{tabular}{|l|c|c|c|c|}
      \hhline{=====}
                                                           & \texttt{Local} & \texttt{Perceptions}   & \texttt{Rendering+App}   & \texttt{Full Offloading} \\ \hline
      GStreamer~\cite{gstreamer}                           & 7              & 13                     & 19                       & 17                       \\
      RaftLib~\cite{beard2017raftlib}                      & 6              & 12                     & 18                       & 16                       \\
      \textbf{\sys}                                        & 5              & 7                      & 9                        & 9                        \\ \hhline{=====}
    \end{tabular}
  }
\end{table}

Table~\ref{tab:reqkernels} shows the number of kernels for GStreamer,
RaftLib, and \sys\ to support AR1 in 
the different scenarios.
GStreamer and RaftLib need auxiliary kernels as local SP libraries.
GStreamer, as a multimedia framework, imposes strict synchronization across streams based on their internal timestamps and requires all kernels to have a single synchronized stream for input and output~\cite{taymans2013gstreamer}.
Since RaftLib does not require such strict 
synchronization, it requires fewer kernels.
GStreamer requires two kernels for branching the camera stream and synchronizing streams for the renderer, while RaftLib only needs a branching kernel.
For the distributed settings, both need additional messaging kernels: 4 for \texttt{Perceptions}, 8 for \texttt{Ren\-der\-ing+App}, and 6 for \texttt{Full Offloading}.
In \sys, these auxiliary kernels are not required because each port can be configured for different usage, which enables the various scenarios flexibly without the system overheads from the auxiliary kernels.

\begin{figure}
  \centering
  \begin{subfigure}[t]{0.24\textwidth}
    \includegraphics[width=\textwidth]{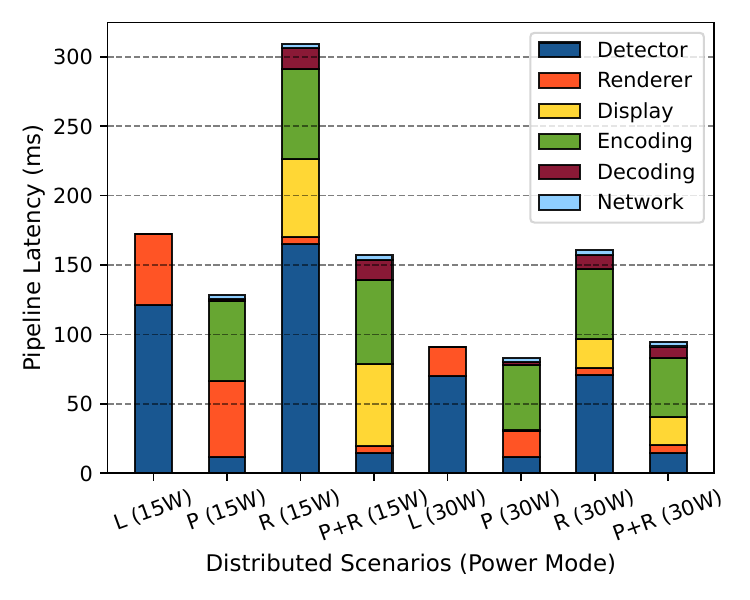}
    \caption{\scriptsize The latency breakdowns of the pipeline components.}
    \label{fig:ar1_pipe_lat}
  \end{subfigure}\hspace{0.008\textwidth}
  \begin{subfigure}[t]{0.175\textwidth}
    \includegraphics[width=\textwidth]{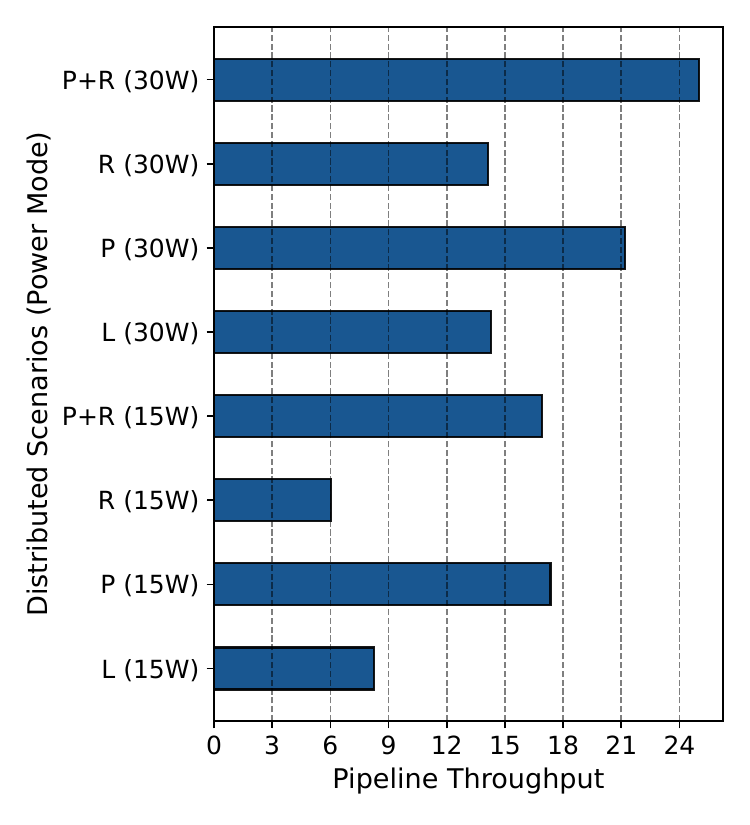}
    \caption{\scriptsize The pipeline throughputs.}
    \label{fig:ar1_pipe_throughput}
  \end{subfigure}
  \vspace{-2.0ex}
  \caption {\small The pipeline latencies and throughputs of AR1 in our distribution scenarios.}
  \label{fig:ar1_res}
   \vspace{-2.0ex}
\end{figure}

\begin{figure}
  \centering
  \begin{subfigure}[t]{0.24\textwidth}
    \includegraphics[width=\textwidth]{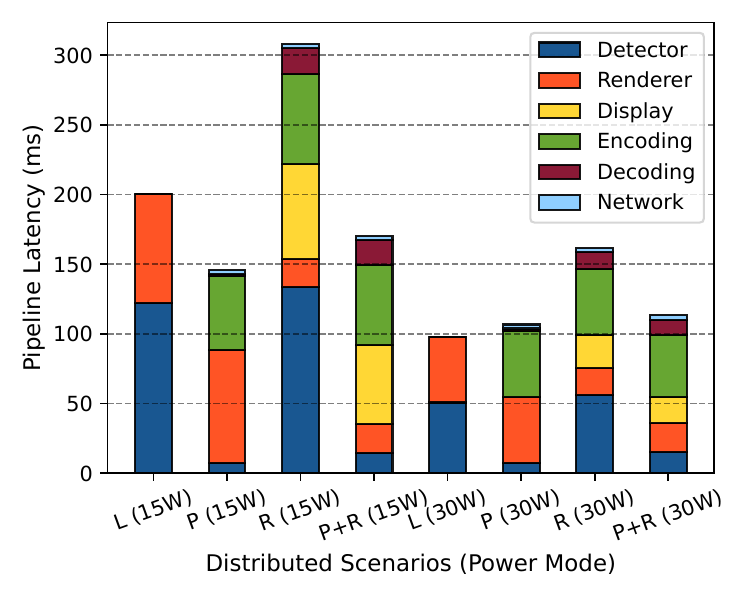}
    \caption{\scriptsize The latency breakdowns of the pipeline components.}
    \label{fig:ar2_pipe_lat}
  \end{subfigure}\hspace{0.008\textwidth}
  \begin{subfigure}[t]{0.175\textwidth}
    \includegraphics[width=\textwidth]{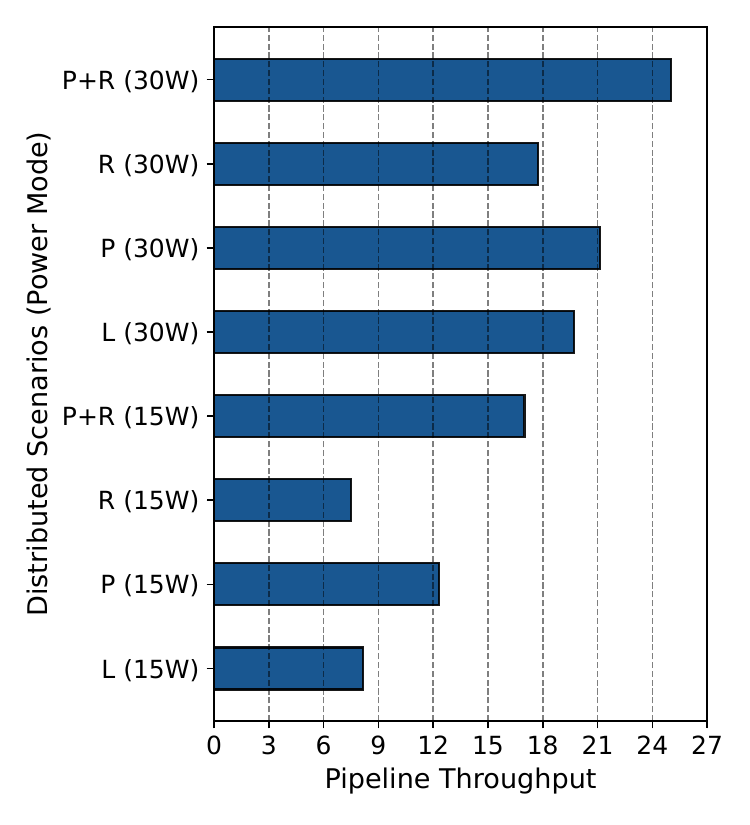}
    \caption{\scriptsize The pipeline throughputs.}
    \label{fig:ar2_pipe_throughput}
  \end{subfigure}
   \vspace{-2.0ex}
  \caption {\small The pipeline latencies and throughputs of AR2 in our distribution scenarios.}
  \label{fig:ar2_res}
   \vspace{-2.0ex}
\end{figure}

\begin{figure}
  \centering
  \begin{subfigure}[t]{0.24\textwidth}
    \includegraphics[width=\textwidth]{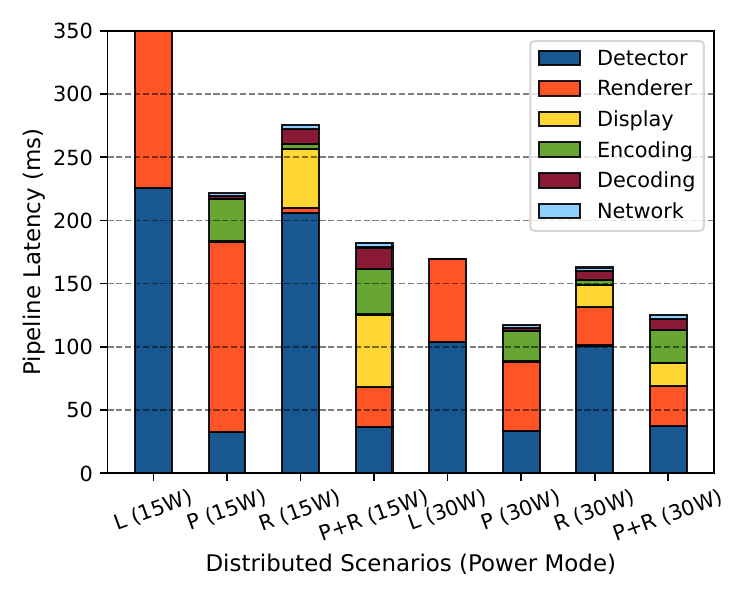}
    \caption{\scriptsize The latency breakdowns of the pipeline components.}
    \label{fig:vr_pipe_lat}
  \end{subfigure}\hspace{0.008\textwidth}
  \begin{subfigure}[t]{0.175\textwidth}
    \includegraphics[width=\textwidth]{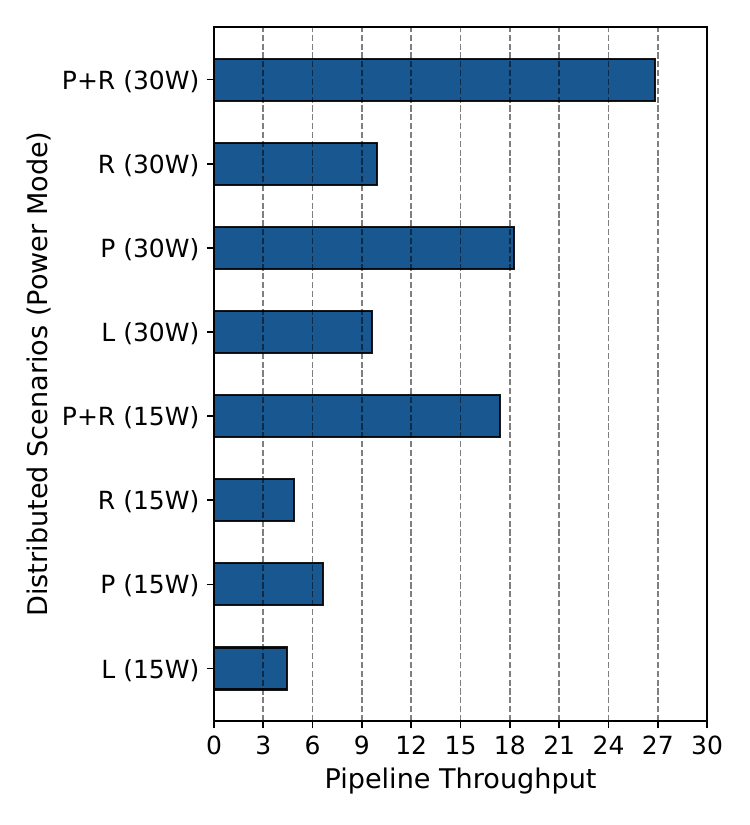}
    \caption{\scriptsize The pipeline throughputs.}
    \label{fig:vr_pipe_throughput}
  \end{subfigure}
   \vspace{-2.0ex}
  \caption {\small The pipeline latencies and throughputs of VR in our distribution scenarios.}
  \label{fig:vr_res}
   \vspace{-2.0ex}
\end{figure}

\subsection{Evaluation of Example Applications}
\label{sec:evalresanaly}
We run the example applications on our testbed of Jetson 15W and 30W with the four distribution scenarios to emulate the situations where the client has little or moderate device capacity.
We measure the average pipeline latency and throughput of the three examples with the scenarios, and demonstrate that the flexibility enabled by \sys\ makes it possible to achieve effective server offloading.

\noindent\textbf{Costs for XR Pipeline Distribution.}
Distributing XR pipelines incurs additional costs: multimedia data compression, displaying the rendered scene from a server, and network transmission.
Transmitting the large multimedia data without compression causes high bandwidth usage with backend network delays and consumes the battery of the user device~\cite{xiao2013modeling, vergara2013energybox}.
Therefore, data (de)compression is necessary for both the client and server.
Another cost is for the client to display the rendered scene from the server.
When the scene is rendered on the server, it should be fetched from GPU memory, sent to the client, and displayed.

The results in Figures~\ref{fig:ar1_res}-\ref{fig:vr_res}, show a breakdown of the average end-to-end latency and average throughput.
The display latency on the client is shown separately from the
rendering latency.
The compression cost is split as encoding and decoding latencies, which include the server- and client-side compression latencies.
The network transmission latency is measured on the client when it receives the result of the timestamped message from the server.

\noindent\textbf{Pipeline Latency and Throughput.}
Figures~\ref{fig:ar1_res}-\ref{fig:vr_res} show the latency and
throughput results of AR1/2 and VR in the distribution scenarios
(\textbf{L}, \textbf{P}, \textbf{R}, and \textbf{P+R}).
Latency is measured as how long the pipeline takes to reflect the real-world
context, and throughput is how frequently the real-world context is reflected to the rendered scene per second.

For AR1 (Figure~\ref{fig:ar1_res}), \textbf{P} shows the lowest latencies in 15W and 30W.
For throughputs, \textbf{P} has the highest throughput in 15W while \textbf{P+R} does in 30W.
Since the pipeline throughput is bound by the dominant functionality, it is possible to have lower throughput even with lower latency.
Compared to \textbf{L}, the throughputs can be improved 2.1\mul\ (15W)
and 1.7\mul\ (30W), and the latencies reduced by 28\% (15W) and 14\% (30W).

For \textbf{P}, the perception on the server takes 11 ms; it takes 121 ms (15W) and 70 ms (30W) of \textbf{L} on the client.
Rendering on the client takes 54 ms (15W) and 19 ms (30W).
\textbf{P} requires the client to encode 1080p camera frames, which takes 57 ms (15W) and 47 ms (30W).
Decoding on the server takes 1.8 ms.
The throughputs of \textbf{P} in 15W and 30W are bound by the encoding latency.

For \textbf{P+R}, since all rendering and perception run on the server, requires  client-side displaying, encoding, and decoding.
On Jet15W, it takes 59 ms for displaying the received FHD scene from the server, 57 ms for encoding camera frames, 12 ms for decoding the received scene.
On Jet30W, it takes 20 ms for displaying, 40 ms for encoding, 6 ms for decoding.
On the server, it takes 14 ms for perception, 5 ms for rendering, 1.7 ms for decoding the received camera frame, and 5 ms for encoding the rendered scene.
So, the throughput of \textbf{P+R} (15W) is bound by client-side displaying while the throughput of \textbf{P+R} (30W) is bound by client encoding.

For AR2 (Figure~\ref{fig:ar2_res} ), \textbf{P+R} shows the highest throughputs while \textbf{P} (15W) and \textbf{L} (30W) present the lowest latencies.
The throughput of \textbf{P+R} is 1.5\mul\ of \textbf{P} (15W) and
1.3\mul\ of \textbf{L} (30W), but it has increase in latencies of 15\% (15W) and 8\% (30W).

For \textbf{P} (15W), the perception takes 7 ms on the server while it does 122 ms of \textbf{L} (15W), and the rendering takes 81 ms.
In addition, it takes 52 ms for client encoding and 1.8 ms for server decoding.
For \textbf{L} (30W), it takes 51 ms for perception and 47 ms for rendering.

For \textbf{P+R}, on Jet15W, it takes 54 ms for encoding the camera, 13 ms for decoding the rendered scene, and 57 ms displaying the received scenes while taking 14 ms for perception and 20 ms rendering on the server.
On Jet30W, it takes 40 ms for encoding, 7 ms for decoding, and 18 ms displaying.
While the throughputs of \textbf{P} (15W) and \textbf{L} (30W) are bound by rendering and perception each, the throughputs of \textbf{P+R} in 15W and 30W are bound by the client encoding.

In Figure~\ref{fig:vr_res} of VR, \textbf{P+R} (15W) and \textbf{P} (30W) present the lowest latencies.
\textbf{P+R} shows the highest throughputs in 15W and 30W.
Compared to \textbf{L}, the throughputs can be 3.9\mul\ (15W) and 2.7\mul\ (30W), and the latencies are 50\% less (15W) and 29\% less (30W).

For \textbf{P}, on Jet30W, it takes 54 ms for rendering, 24 ms for encoding 480p camera frames.
On the server, it takes 33 ms for perception and 1.5 ms for decoding the received camera frames.
On Jet15W, it takes 150 ms for rendering and 33 ms for encoding.
The throughputs of \textbf{P} in 15W and 30W are bound by the rendering.
Since the rendering is so challenging for Jet15W, it dominates throughput and latency.

For \textbf{P+R}, on Jet15W, it takes 57 ms for displaying the scene from the server, 31 ms for encoding camera frames, 15 ms for decoding the received scene.
On Jet30W, it takes 18 ms for displaying, 20 ms for encoding, 7 ms for decoding.
On the server, it takes 36 ms for perception, 31 ms for rendering, 1.7 ms for decoding the received camera frame, and 5 ms for encoding the rendered scene.

For Jet15W, rendering and perception are challenging, and \textbf{P+R} is beneficial in terms of latency and throughput.
In the case of Jet30W, even though \textbf{P+R} introduces additional overheads for the client-side decoding and displaying, the pipeline bottleneck of rendering is relieved compared to \textbf{P} (30W), enabling higher throughput.

\noindent\textbf{Result Analysis.}
In the results in Figure~\ref{fig:ar1_res}, the optimal distribution scenario can vary in terms of the latency and throughput even with the same workloads by the client capacity: \textbf{P} and \textbf{P+R} in 15W and 30W.
For higher throughput, it is crucial to offload the pipeline bottleneck.
There are cases where the distribution overheads are larger than the benefits, ending up with worse performance than the local-only scenario (\eg, \textbf{R} (15W) and (30W) of AR1 and AR2 in Figure~\ref{fig:ar1_res} and~\ref{fig:ar2_res}).
The server and client device capacities should be considered because the kernels are parallelized and can cause resource contention.
For instance, the perception latencies of \textbf{P+R} in AR1, AR2, and VR increase on the server compared to \textbf{P}.
Moreover, although AR1 and AR2 are with the same pipeline structure of
Figure~\ref{fig:arconfigs}, the ideal distribution is different based on the perception and rendering complexities of each application.

Based on our results, the effectiveness of offloading depends on the given workloads, server and client capacities, and offloading overheads.
\sys\ allows each user to configure the workload distribution flexibly at runtime and enables the optimal distribution of an XR application for various distribution scenarios.

\section{Discussion}
\label{sec:discussion}

FleXR offers flexibility in how an XR workload is distributed across a device and offload server(s). This opens up several opportunities for innovation and new research directions.

First, while with \sys\ the XR configuration can be tuned to the specifics of the deployment context to realize optimal workload distributions, currently, this requires manual effort from the system users.
To fully realize the potential of \sys, future research is needed on automated deployment and resource management.
New methods are needed to consider factors such as kernel costs, client and server capacity, network state, and offloading overheads, as well as to enable dynamic adaptation of the workload configurations.
Previous function offloading systems have used linear solvers and resource profilers for offloading decisions with static analysis~\cite{chun2011clonecloud, cuervo2010maui}, but this approach would be more complex in \sys\ as the kernel costs and offloading overheads are subject to change based on user and server situations.

Second, by making it possible to integrate third-party components and application frameworks with \sys\ (\eg, the game engines), we make it possible to consider a future landscape of XR supported by distributed edge-cloud infrastructure, potentially with different performance, quality, or other properties, that can be combined in different ways to support complex future XR use cases. This new landscape opens up new challenges for distributed orchestration for XR, and also promotes the reuse of service functionality in different scenarios, thus enabling faster innovation.

Finally, while distributed service composition has been considered in other contexts~\cite{servicecomposition1,servicecomposition2}, several XR-specific aspects raise new challenges and opportunities that future work should address. These are related to performance/timeliness and quality tradeoffs that exist at the application level, sharing and reuse across users (\eg, as done for specific offload services in~\cite{ran2019sharear}), XR-specific transport protocols~\cite{braud2017future, abdallah2018delay}, new privacy concerns, \etc\ By creating and open sourcing the \sys\ infrastructure, we believe our work will facilitate such research directions.

\section{Conclusion}
\label{sec:conclusion}
In this work, we describe the limitations of the existing distributed XR systems by their predetermined architectures.
We argue the need of flexibility in XR workload distribution, and the lack of flexibility in distributed XR is attributed to the absence of system supports.
For enabling such flexibility, we present \sys\ -- a DSP system enabling the flexible distribution of XR pipelines at runtime.
We identify the issues of the DSP design for XR and resolve them while building \sys.
In our experiments, \sys\ efficiently enables the four different distribution scenarios to three XR use cases.
Our results show the optimal workload distribution is relatively determined by environmental factors, and \sys\ can realize the effective server assistance for each XR use case in various environments.

\section*{Acknowledgment}
\label{sec:ack}
We would like to thank the anonymous reviewers.
This work has been partially supported by NSF projects CCF-2217070 and CNS-1909769, the Applications Driving Architectures (ADA) Research
Center, a JUMP Center co-sponsored by SRC and DARPA, and by funding and equipment gifts from VMware and Intel.

\bibliographystyle{ACM-Reference-Format}
\bibliography{sample-base}

\end{document}